\documentclass[prl,twocolumn,preprintnumbers,amsmath,amssymb]{revtex4}
\usepackage{dcolumn}
\usepackage{bm}
\usepackage{comment}
\usepackage{amsmath}
\usepackage{amssymb}
\usepackage{amsfonts}
\usepackage{eucal}
\usepackage{color}
\usepackage[active]{srcltx}%

\newcommand{\pd}{\partial}

\def\nn{\nonumber}

\def\nhegalgebra{\widehat{\mathcal{V}_{\vec{k},S}}}

\newcommand{\ba}{\begin{array}}
\newcommand{\ea}{\end{array}}
\newcommand{\be}{\begin{equation}}
\newcommand{\ee}{\end{equation}}
\newcommand{\bea}{\begin{eqnarray}}
\newcommand{\eea}{\end{eqnarray}}
\newcommand{\bse}{\begin{subequations}}
\newcommand{\ese}{\end{subequations}}
\newcommand{\bi}{\begin{itemize}}
\newcommand{\ei}{\end{itemize}}

\newcommand{\eps}{\epsilon}
\newcommand{\p}{\partial}

\definecolor{darkgreen}{rgb}{0,0.3,0}
\definecolor{darkblue}{rgb}{0,0,0.3}
\definecolor{darkred}{rgb}{0.7,0,0}

\begin{document}

\title{
Extremal Rotating Black Holes in the Near-Horizon Limit:\\ Phase Space and Symmetry Algebra}

\author{G. Comp\`{e}re$^\dag$, K. Hajian$^{\ddag, \S}$, A. Seraj$^{\ddag}$, M.M. Sheikh-Jabbari$^\ddag$ }

\affiliation{\vspace*{3mm}$^\dag$ \textit{Universit\'{e} Libre de Bruxelles and International Solvay Institutes
CP 231 B-1050 Brussels, Belgium
}
\vspace*{2mm}\\ $^\ddag$ \textit{School of Physics, Institute for Research in Fundamental
Sciences (IPM), P.O.Box 19395-5531, Tehran, Iran} \vspace*{2mm}\\ $^\S$
\textit{Department of Physics, Sharif University of Technology,
 P.O. Box 11365-8639, Tehran, Iran}}
\vfil

\begin{abstract}
\noindent We construct the \emph{NHEG phase space}, the classical phase space of Near-Horizon Extremal Geometries with fixed angular momenta and entropy, and with the largest symmetry algebra. We focus on vacuum solutions to $d$ dimensional Einstein gravity. Each element in the phase space is a geometry  with $SL(2,\mathbb R)\times U(1)^{d-3}$ isometries which has vanishing $SL(2,\mathbb R)$  and constant $U(1)$ charges.
We construct an on-shell vanishing symplectic structure, which leads to an infinite set of symplectic symmetries.
In four spacetime dimensions, the phase space is unique and the symmetry algebra  consists of the familiar  Virasoro algebra, while in $d > 4$ dimensions the symmetry algebra, the \emph{NHEG algebra},  contains infinitely many  Virasoro subalgebras. The nontrivial central term of the  algebra is proportional to the black hole entropy. The conserved charges {are given by} the Fourier decomposition of a Liouville-type stress-tensor which depends upon a single periodic function of $d-3$ angular variables associated with the $U(1)$ isometries. This phase space and in particular its symmetries can serve as a basis for a semiclassical description of extremal rotating black hole microstates.
\end{abstract}


\maketitle

Questions regarding black holes have been at the frontiers of  astrophysics and high energy physics. On the theoretical side the possible microscopic origin of thermodynamical aspects of black holes \cite{BCH},  the  information loss problem and its the recent developments \cite{Information-Loss},   have been  active research areas in the last forty years. These questions are usually regarded as test grounds for, and windows to,  models of quantum gravity. On the observational side, and with the advance in X-ray astronomy (see e.g. \cite{X-ray-Chandra}), we now have several approved candidates of black holes in a wide range of masses and spins.
Extremal spinning black holes, namely black holes with maximum possible spin for a given mass, are an important special class of black holes to study.  Remarkably, several near-extremal Kerr black holes have been observationally identified \cite{Extreme-Kerr-observation}. In the extremal limit, the Hawking temperature vanishes and very close to the horizon one finds a Near-Horizon Extremal Geometry (NHEG) with  enhanced $SL(2,\mathbb R)\times U(1)$ isometry where the dynamics is  decoupled from the region far from the black hole horizon \cite{Bardeen-Horowitz}. The Kerr NHEG can therefore be  an appealing starting point for analytic modeling  of physical phenomena 
around astrophysical near-extreme rotating black holes. 

Earlier analyses have established uniqueness of the Kerr NHEG as the 4d Einstein vacuum solution with $SL(2,\mathbb R)\times U(1)$ isometry \cite{KLR}. This uniqueness  has been extended to more general {solutions} to pure Einstein vacuum {gravity (with or without cosmological constant)} in $d$ dimensions with $SL(2,\mathbb R)\times U(1)^{d-3}$ isometry \cite{NHEG-review}. The latter is the class of solutions we focus on in this work. The metric has the general form
\be\label{NHEG-metric}
\overline{ds}^2=\Gamma(\theta)\left[ds^2_2+d\theta^2+\gamma_{ij}(\theta)(d\varphi^i+k^irdt)(d\varphi^j+k^jrdt)\right]
\ee
where $ds^2_2=-r^2dt^2+\frac{dr^2}{r^2}$ and $i,j=1,2,\cdots, d-3$. We require the geometry to be smooth and Lorentzian. The latter implies  $\Gamma>0$ and the eigenvalues of $\gamma_{ij}$ to be nonnegative. {We work with Poincar\'e coordinates for $AdS_2$ since these coordinates appear naturally in the near-horizon limit and are preferred to match the region outside the near-horizon region. Our results, as we discuss, are independent of this choice.}

The solution \eqref{NHEG-metric} is specified by $d\!-\!3$ constant parameters $\vec{k} = (k^1,\dots , k^{d-3})$ which are thermodynamically conjugate to angular momenta $\vec{J}$. One can associate an entropy $S$ to this geometry which is a Noether-Wald \cite{Iyer-Wald} conserved charge \cite{HSS1} and obeys the entropy law \cite{HSS1,Astefanesei:2006dd} 
\be\label{Entropy-Law}
\frac{S}{2\pi}\equiv \frac{1}{8 \pi G}\oint_\mathcal{H} {d\theta \, d\vec{\varphi}\, }\Gamma^{\frac{d-2}{2}}\sqrt{\det\gamma} = \vec{k}\cdot \vec{J}\,.
\ee
Here,  $\mathcal{H}$ denotes codimension two, constant arbitrary $t,r$ surfaces. Such $\mathcal{H}$'s form infinitely many bifurcation surfaces of  the geometry \eqref{NHEG-metric}, as detailed in \cite{HSS1, Longer-paper}. 

{There have been many proposals for understanding the possible microscopic origin of the extremal black hole entropy. One can recognize two classes of such proposals. In the top-down approach, the extremal black hole is embedded into a consistent quantum gravity such as string theory. Microstates of some classes of supersymmetric black holes can then be counted microscopically, see e.g. \cite{Strominger-Vafa,Sen}. 
In the bottom-up approach, one builds  upon classical and semiclassical properties of not necessarily supersymmetric black holes and then infer a possible holographic theory, inspired by the AdS/CFT correspondence \cite{Maldacena:1997re}, which allows to effectively count the number of microstates, see e.g. \cite{Carlip,Kerr/CFT}. Such an approach relies on the appearance of an AdS$_2$ factor in the near-horizon region and benefits from the universality of the attractor mechanism \cite{Sen:2005wa}.}

In this paper, we introduce the framework  of a new kind of bottom-up proposal. We  construct the  NHEG phase space: the set of all geometries which are diffeomorphic to, but physically distinct from, \eqref{NHEG-metric}. The distinction comes from conserved charges associated with each geometry in the phase space. The geometries in the phase space fall into representation of the NHEG algebra, 
 the symmetry of the phase space realized as the  Dirac bracket of the associated conserved charges. The symmetry algebra admits a central charge which is the black hole entropy. The existence of a symplectic structure, which we explicitly construct here, allows for a semiclassical quantization of the phase space. Here, we summarize our results   while details of the analysis will be given in \cite{Longer-paper}. We also comment on the quantization of the phase space and the relationship with the Kerr/CFT proposal \cite{Kerr/CFT} in the discussion section.

\vskip 2mm

\noindent\textbf{Summary of the results:}
\vskip 1mm

\textbf{\emph{The NHEG phase space.}} Our main motivation for considering diffeomorphisms as the basis for the construction of our phase space comes from the absence of dynamical physical perturbations around the background as explicitly shown for vacuum four dimensional Einstein gravity in \cite{No-Dynamics}. {Since the main arguments of \cite{No-Dynamics} rely on the existence of an $AdS_2$ factor which appears in any dimension, we expect that these arguments extend to generic NHEG backgrounds. Moreover, assuming that perturbations are invariant under the 2d subgroup of $SL(2,\mathbb{R}$) it was proved in \cite{HSS2} that the ``no dynamics" argument extends to generic near horizon extremal geometries which admit a background uniqueness theorem \cite{NHEG-review}.} Therefore, 
 we are naturally led to construct the (semi)classical phase space of near-horizon extremal geometries with given angular momenta by the action of diffeomorphisms on \eqref{NHEG-metric}. The vector field which, as we will outline, is appropriate for this purpose is within the family $\chi[\epsilon(\vec{\varphi})]$
\begin{align}\label{ASK}
\chi[{\epsilon}(\vec{\varphi})]&={\epsilon}(\vec{\varphi})\vec{k}\cdot\vec{\pd}_\varphi-\vec{k}\cdot\vec{\pd}_{\varphi}{\epsilon}\;(\dfrac{1}{r}\pd_{t}+r\pd_{r}),
\end{align}
where $\epsilon(\vec{\varphi})$ is an arbitrary periodic function of {$\varphi^1,\dots \varphi^{d-3}$. Under the $x^\mu\to x^\mu- \chi^\mu$ diffeomorphisms, metric \eqref{NHEG-metric} changes as $g_{\mu\nu}\to g_{\mu\nu}+{\cal L}_\chi g_{\mu\nu}$, where  ${\cal L}_\chi$ is the Lie derivative along $\chi$. The finite coordinate transformation built from \eqref{ASK} is $\bar{x}^\mu \to x^\mu$ where
\begin{align}\label{Finite-Diffs}
\begin{split}
\bar{\varphi}^i&=\varphi^i + k^i F(\vec{\varphi}), \qquad \qquad \bar{\theta}=\theta,\\
\bar{r} &=re^{-{\Psi(\vec{\varphi})}},\qquad \bar{t} =t-\frac{(e^{\Psi(\vec{\varphi})}-1)}{r},
\end{split}
\end{align}
and $\Psi$ is defined through
\be\label{Phi-def}
e^{\Psi}=1+\vec{k}\cdot{\vec{\pd}}_{\varphi } F(\vec{\varphi}).
\ee
{With $F(\vec{\varphi}) = \eps(\vec\varphi)$ infinitesimal, one recovers the infinitesimal diffeomorphism \eqref{ASK}.}

 With the above we construct the phase space $\mathcal{G}[\{F\}]$ as the family of metrics obtained through \eqref{Finite-Diffs}, {viewed as an active transformation}.  $\mathcal{G}[\{F\}]$ is the collection of all metrics with arbitrary periodic function $F(\vec{\varphi})$ explicitly given by
\begin{align}\label{NHEG-phase-space}
 ds^2=\Gamma(\theta)&\Big[-\left( \boldsymbol\sigma -  d\Psi \right)^2+\Big(\dfrac{dr}{r}-d{\Psi}\Big)^2
\nn \\ &+d\theta^2+\gamma_{ij}(d\tilde{\varphi}^i+{k^i}\boldsymbol{\sigma})(d\tilde{\varphi}^j+{k^j}\boldsymbol{\sigma})\Big],
\end{align}
where $\tau=t+\frac{1}{r}$ and
\be
\boldsymbol{\sigma}=e^{-{\Psi}}rd\tau+\dfrac{dr}{r},\qquad \tilde{\varphi}^i=\varphi^i+k^i (F-{\Psi})\,.\nonumber
\ee
The background \eqref{NHEG-metric} is the $F=0$ element in $\mathcal{G}[\{F\}]$. Obtained from diffeomorphisms \eqref{Finite-Diffs}, $\mathcal{G}[\{F\}]$ contains metrics which are smooth everywhere. 
We will be defining the conserved charges through integration of $(d-2)$-forms on the constant $t,r$ surfaces ${\cal H}$ which are bifurcation surfaces of Killing horizons of NHEG geometry \cite{HSS1,Longer-paper}\footnote{We note that the Killing horizons of the NHEG geometry should not be confused with the Killing horizon of the extremal black hole whose near horizon limit leads to the NHEG. In particular note that the NHEG has infinitely many bifurcate Killing horizons \cite{HSS1, Longer-paper}, while the horizon of any extremal black hole is degenerate and non-bifurcate.}. An interesting property of the phase space $\mathcal{G}[\{F\}]$ is that the induced metric on surfaces $\cal H$ is smooth and has the same form for any constant $t,r$ surface and for any configuration of the phase space,
\begin{align}
ds^2_{\cal H}=\Gamma(\theta)\left[d\theta^2+\gamma_{ij}(\theta)\,d\tilde{\varphi}^i\, d\tilde{\varphi}^j\right].
\end{align}
{Given our construction above, one clearly sees that the $SL(2,\mathbb R)\times U(1)^{d-3}$ isometries of the background extend to each metric of the form \eqref{NHEG-phase-space} in the phase space $\mathcal{G}[\{F\}]$. Notice that the angular momenta are not associated with $\p_{\varphi_i}$ but rather with the background $U(1)$ Killing vector fields transformed by the diffeomorphism \eqref{Finite-Diffs} \cite{Longer-paper}. This implies that the angular momenta, defined as Komar integrals, are constant over the phase space. Also,  each bifurcate Killing horizon has a bifurcation surface with the same area as the background. In that sense, the phase space {contains} geometries of equal entropy $S$ and angular momenta $\vec{J}$.

{
The most important property of the NHEG phase space is the existence of a finite and conserved symplectic structure,  allowing one to define the classical and semiclassical dynamics. The standard Lee-Wald symplectic structure \cite{Lee-Wald} built from the Einstein action diverges, as was noted in \cite{AMR}. Nonetheless, as we will discuss below, there exist boundary terms which once added remove the divergences. The resulting symplectic form vanishes everywhere on-shell. In the analogous case of vacuum Einstein gravity in three dimensions, there is also no bulk dynamics while boundary conditions exist which enjoy two copies of the Virasoro algebra as symmetry algebra \cite{Brown-Henneaux}. In that setting, it has been recently shown in \cite{Compere:2014cna} that the symplectic form vanishes on-shell on the phase space \cite{Banados}, which implies that the symmetries act everywhere in the bulk spacetime. The situation is analogous here. Since the symplectic form is zero on-shell instead of at infinity only, the asymptotics is not a special place and symmetries act everywhere. We will hence refer to them as {\it symplectic symmetries} in contrast with asymptotic symmetries.

\vskip 2mm
\textbf{\emph{The NHEG symplectic symmetry algebra.}} Since the symplectic structure is nontrivial off-shell, one can define physical surface charges associated with the symplectic symmetries $\chi[\eps_{\vec{n}}]$, where $\eps_{\vec{n}}=e^{i \vec{n} \cdot \vec{\varphi}}$, $n_i\in \mathbb{Z}$. The generators of these charges is denoted by $L_{\vec{n}}$.  As is standard practice; e.g. see \cite{Compere-thesis}, once given the symplectic structure one can read off the classical algebra of charges and the corresponding central charge. This  algebra can then be quantized by replacing the classical bracket by $-i\hbar$ times the commutator. We hence obtain the quantum algebra of charges, the NHEG algebra $\nhegalgebra$:
\bea\label{NHEG-algebra}
[ L_{\vec{m}}, L_{\vec{n}} ] = \vec{k} \cdot (\vec{m}- \vec{n}) L_{\vec{m}+\vec{n}} +  \frac{S}{2\pi}(\vec{k}\cdot \vec{m})^3 \delta_{\vec{m}+\vec{n},0}\,.
\eea
The angular momenta  $J_i$ and the entropy $S$ obeying \eqref{Entropy-Law} commute with $L_{\vec{n}}$ and are therefore central elements of the algebra. The full symmetry algebra of {the} semiclassical phase space is then
\be\label{Agebra-Full}
SL(2,\mathbb{R})\times U(1)^{d-3}\times \nhegalgebra .
\ee

For the four dimensional {Kerr case},  $k=1$ and one obtains the familiar Virasoro algebra 
\bea\label{Virasoro}
[ L_m, L_n  ] = (m-n) L_{m+n} + \frac{c}{12}m^3\delta_{m+n,0}
\eea
with central charge $c= 12 \frac{S}{2 \pi}=12J$, which is the same algebra appearing in Kerr/CFT setup \cite{Kerr/CFT}. Note that despite the similarity, as we will discuss further  at the end of this Letter, our construction has crucial conceptual and technical differences with Kerr/CFT.

In higher dimensions, the NHEG algebra \eqref{NHEG-algebra} is a new infinite-dimensional algebra in which the entropy appears as the central extension. For $d>4$ the algebra contains infinitely many Virasoro subalgebras. {To see the latter, one may focus on the generators $L_{\vec{n}}$ where $\vec{n}=n \vec{e}$ for any given vector on the lattice $\vec{e}$, $\vec{e}\cdot \vec{k}\neq 0$.} It is then readily seen that $\ell_n\equiv \frac{1}{\vec{k}\cdot \vec{e}} L_{\vec{n}}$ form a Virasoro algebra of the form \eqref{Virasoro} with central extension $c=\frac{12S}{2\pi}\vec{k}\cdot \vec{e}$. The entropy might then be written in the suggestive form $S= \frac{\pi^2}{3} c \,T_{F.T.}$ where $T_{F.T.}^{-1} = 2\pi\vec{k}\cdot \vec{e}$ is the inverse Frolov-Thorne temperature, as reviewed in \cite{Compere-review}. The algebra also contains many infinite dimensional Abelian subalgebras spanned by generators of the form $L_{\vec{n}}$ where $\vec{n} = n \vec{v} $ and $\vec{v} \cdot \vec{k} = 0$, under the condition that $\vec{v}$ is on the lattice.

\vskip 2mm
\textbf{\emph{On the choice of symmetry {generator}.}} The background \eqref{NHEG-metric} enjoys $SL(2,\mathbb R)\times U(1)^{d-3}$ isometry. Let us denote the  $SL(2,\mathbb R)$ generators by $\overline\xi_{-}, \overline\xi_0,\overline\xi_+$
\be\label{xi1-xi2}\begin{split}
\overline\xi_- &=\partial_t\,,\qquad\overline \xi_0=t\partial_t-r\partial_r,\\
\overline\xi_+ =\dfrac{1}{2}&(t^2+\frac{1}{r^2})\partial_t-tr\partial_r-\frac{1}{r}\vec{k}\cdot \vec{\p}_\varphi.
\end{split}
\ee
We also define the two vectors
\begin{equation}\label{eta1-eta2}
\overline\eta_1 =\frac{1}{r}\p_t\,, \quad \overline\eta_2=r\p_r\,,
\end{equation}
{and denote by $\xi_-,\xi_0,\xi_+,\eta_1,\eta_2$ the push-forward of these vectors on a generic element of the phase space after acting with the diffeomorphism \eqref{Finite-Diffs}.}
Starting with the most general diffeomorphism generator $\chi$, we highlight  conditions singling out {\eqref{ASK}}, which is the basic object both in construction of the phase space $\mathcal{G}[\{F\}]$ and the algebra \eqref{NHEG-algebra}. The following six requirements  \emph{uniquely} fix $\chi$ given in \eqref{ASK}. {These requirements are mainly aimed at providing a rationale for selecting the diffeomorphism which was found by an ansatz.}
\begin{enumerate}
\item $[\chi,\overline\xi_{-}]=0=[\chi,\overline\xi_{0}]$. This condition implies
$$
\chi=\frac{1}{r}\eps^t\p_t+r\eps^r\p_r+\eps^\theta\p_\theta+\vec{\epsilon}\cdot\vec{\p}_{\varphi},
$$
where all components are functions of $\theta,\vec{\varphi}$. This implies that $\xi_- = \overline \xi_-$ and $\xi_0 = \overline \xi_0$ are Killing isometries of each element of the phase space $\mathcal{G}[\{F\}]$.

An arbitrary $t,r$ can be mapped onto any given constant $t_0,r_0$ under a $\xi_-,\xi_0$ transformation. $\xi_-,\xi_0$ invariance implies that the charges associated with geometries in the NHEG phase space $\mathcal{G}[\{F\}]$ are independent of  the codimension two surface $\cal{H}$ (bifurcation horizons of the NHEG) over which the charges are defined.

We also comment that $\overline\eta_a$ are $\xi_-,\xi_0$ invariant; i.e. $[\overline\eta_a,\xi_b]=0, \ a=1,2,\ b=-1,0$.
\item $\nabla_\mu \chi^\mu=0$ and hence the volume element $\boldsymbol{\epsilon}$,
\begin{equation}
\boldsymbol{\epsilon}=\frac{1}{d!}\sqrt{-g}{\epsilon}_{\mu_1\mu_2\cdots\mu_d}dx^{\mu_1}\wedge dx^{\mu_2}\wedge\cdots \wedge dx^{\mu_d},
\end{equation}
is the same for all elements in $\mathcal{G}[\{F\}]$, i.e.  $\delta_{\chi}\boldsymbol{\epsilon}=0$.
\item {$\delta_\chi\mathbf{L}=0$,} where $\mathbf{L}=\frac{1}{16\pi G} R\boldsymbol{\epsilon}$ is the Einstein-Hilbert  Lagrangian $d$-form computed over the background \emph{ansatz} \eqref{NHEG-metric} before imposing the equations of motion. The above two properties lead to $\eps^\theta=0$ and
$
\eps^{r}=
 -\vec{\partial}_{\varphi}\cdot \vec{\epsilon}\,.$
\item We fix $\eps^t = - b \, \vec{\partial}_{\varphi}\cdot \vec{\epsilon}$. {Upon further imposing $b=1$,} the diffeomorphism then preserves one of two expansion-free rotation-free and shear-free null geodesic congruences which is labelled by the normal to constant $v = t+\frac{1}{r}$ surfaces (The other congruence is related to $u=t-\frac{1}{r}$) \cite{Longer-paper, Durkee:2010ea}.

\item We impose  $\vec{\eps}$ to be $\theta$ independent. {This condition along with condition 4 above lead to
\begin{align}
\chi_b[{\epsilon}(\vec{\varphi})]={\epsilon}(\vec{\varphi})\vec{k}\cdot\vec{\pd}_\varphi-\vec{k}\cdot\vec{\pd}_{\varphi}{\epsilon}\;(\dfrac{b}{r}\pd_{t}+r\pd_{r}).\nonumber
\end{align}
Let us study the smoothness of the $t,r$ constant surfaces ${\cal H}$. For a generic choice of $b$ we would have 
	\begin{align}\label{H-surface}
	\qquad ds^2_{\cal H}=\Gamma(\theta)\left[(1-b^2)d{\Psi}^2+d\theta^2+\gamma_{ij}(\theta)\,d\tilde{\varphi}^i\, d\tilde{\varphi}^j\right].
	\end{align}
The first term violates the smoothness of $\cal H$ at poles unless $b=\pm 1$. We kept the dependence in $b$ to demonstrate that the choice $b=0$ which was used in \cite{Kerr/CFT} leads to a lack of smoothness of $\mathcal H$. (Moreover, this choice does not preserve one of the special geodesic congruences.) We take $b=1$ from now on. Note that the lack of $\theta$ dependence also makes the volume of ${\cal H}$ be invariant under $\chi$-diffeomorphisms, as is explicit from \eqref{H-surface} after checking $\tilde\varphi^i \sim \tilde\varphi^i +2\pi$, which leads to a conserved entropy.
}

\item  We require finiteness, conservation and regularity of the symplectic structure. This leads  to $\vec{\eps}= \vec{v} \eps$
where $\vec{v}$ is a constant fixed direction. If $\vec{v}$ is along $\vec{k}$ the function $\eps$ can be a function of all coordinates $\vec{\varphi}$, otherwise it can be only a function of the coordinate along $\vec{v}$. That is, we have two families of generators: (i) $\vec{\eps} \cdot \p_{\vec{\varphi}} = \eps(\phi)\p_\phi$ where $\phi$ is a specific $SL(d-3,\mathbb Z)$ choice of circle in the $(d\!-\!3)$-torus spanned by $\vec{\varphi}$; (ii) $\vec{\eps} = \vec{k} \eps(\vec{\varphi} )$.

The first choice leads to a {family of ``Kerr/CFT phase spaces''}, that we will discuss in \cite{Longer-paper}. The second choice leads to the NHEG phase space $\mathcal{G}[\{F\}]$ that we describe here.

\end{enumerate}

\vskip 2mm
\textbf{\emph{The symplectic structure.}}
The solution space $\mathcal{G}[\{F\}]$ can be promoted to a phase space only when the symplectic structure is defined.
It is well-known that the Lee-Wald $(d\!-\!1)$ symplectic form $\boldsymbol \omega_{LW}[\delta_1\Phi,\delta_2\Phi;\Phi] $ for a generic theory with fields $\Phi$ and field variations $\delta\Phi$  is ambiguous up to the addition of  boundary terms  \cite{Lee-Wald}. According to the holographic renormalization framework, the total symplectic form takes the form
\be\label{omega-total}
\hspace*{-3mm}\boldsymbol \omega[\delta_1\Phi,\delta_2\Phi;\Phi] = \boldsymbol \omega_{LW} + \boldsymbol  d (\delta_1 \boldsymbol Y[\delta_2\Phi,\Phi] - \delta_2 \boldsymbol Y[\delta_1\Phi,\Phi]),
\ee
where $\boldsymbol Y[\delta\Phi,\Phi]$ is the $(d\!-\!2)$-form boundary pre-symplectic potential \cite{Compere-Marolf}. The symplectic structure is then defined for a codimension one surface $\Sigma$ as $\int_\Sigma  \boldsymbol \omega $.
 Since we only consider diffeomorphisms, metric variations are Lie derivatives, $\delta_\chi g_{\mu\nu}={\cal L}_\chi g_{\mu\nu}$.

We fix the ansatz for $\boldsymbol Y[\delta\Phi,\Phi]$
by requiring the following. (a) Since the bulk action has two derivatives, we require $\boldsymbol Y$ to have at most one derivative. (b) We allow $\boldsymbol Y$ to depend on the metric and on $\eta_1$, $\eta_2$. 
We then restrict the corresponding coefficients through the following requirements: (i) The symplectic structure should be finite and conserved. Given the $\xi_-,\xi_0$ invariance,  one has $\omega^t \sim 1/r$, $\omega^r \sim r$. This leads to a logarithmically divergent symplectic structure with infinite flux unless $\omega^t =0= \omega^r$ on-shell, which we therefore require. {(ii) We require that $\omega^\theta = 0 = \omega^{\varphi^i}$ on-shell. It implies that any smooth deformation of the surface $\mathcal H$ will lead to the same conserved charges. {(iii) We require that the central charge should be independent on $b$.} We find that a boundary term which guarantees these requirements is
\bea\label{boundary-term}
\boldsymbol Y = -i_{{\eta_1+\eta_2}} \cdot \boldsymbol \Theta + \frac{1}{{16 \pi G}}
{(\eta_1^\alpha+\eta_2^\alpha)} \delta g_{\alpha\beta}\eta_1^\beta\star \boldsymbol \epsilon_\perp
\eea
where  $\boldsymbol\Theta[\delta g_{\mu\nu} , g_{\mu\nu}] $ is the $d-1$ form appearing in the on-shell variation of the Einstein action $\delta {\mathbf{L}}\approx d\boldsymbol{\Theta}$ \cite{Iyer-Wald} and $\boldsymbol \epsilon_\perp$ is the binormal to the two shear-free expansion-free and rotation-free null congruences, normalized as  $\boldsymbol \epsilon_\perp=dt\wedge dr$ on the background. No boundary term in the class exists when $\vec{\eps} = \vec{K} \eps(\varphi^i)$, with $\eps$ an arbitrary function of all angles $\varphi^i$ and $\vec{K} \neq \vec{k}$, which justifies the { last} requirement in the choice of symmetry generator.

\vskip 2mm
\textbf{\emph{Integrability condition.}}
Given the symplectic form $\boldsymbol \omega$, we can define variations of surface charges
around any element of the phase space \eqref{NHEG-phase-space}. One consistency requirement is to be able to integrate these charge variations into finite charges. The latter is known as the \emph{integrability conditions} which read as \cite{Wald-Zoupas}
$
\int_{\cal H} \chi \cdot \boldsymbol \omega[\delta_1 \Phi,\delta_2 \Phi; \Phi] =0
$
for any field variations $\delta_1 \Phi$, $\delta_2 \Phi$ and fields $\Phi$ and any symmetry generator $\chi$. In our case the integrability conditions are obeyed as a consequence of $\chi^t \omega^r=\chi^r \omega^t$ which holds off-shell.

\textbf{\emph{The conserved charges.}}
Given the symplectic structure one can compute the charges $Q_\chi$ \cite{Lee-Wald}. To this end one may start from  the fact that charge variations are defined through the Poisson bracket of charges, $\delta_{\chi_2}Q_{\chi_1}=\{Q_{\chi_1},Q_{\chi_2}\}=Q_{\{\chi_1,\chi_2\}}+C(\chi_1,\chi_2)$, where $C$ is the central element, and then deduce the charges $Q_\chi$. 
It is straightforward to check that acting on the phase space with the symmetry generator $\chi[\eps(\vec{\varphi})]$, keeps the metric in the same functional form as \eqref{NHEG-phase-space} but with $F$ shifted as $\delta_{\eps} F=  (1 + \partial F )\eps=e^{{\Psi}}\eps$ where $\partial$  denotes the ``directional derivative'' $\p \equiv \vec{k} \cdot \vec{\p}$. One can translate this transformation law in terms of  ${\Psi}$ defined in \eqref{Phi-def} as
\bea
\delta_\eps {\Psi} =\eps \p {\Psi}+  \p \eps.
\eea
Therefore ${\Psi}$ transforms like a Liouville field, which we dub as the {\it NHEG boson} and
\bea\label{T-tensor}
T [\Psi]= \frac{1}{16\pi G}\left( (\p {\Psi})^2- 2 \p^2 {\Psi}+2e^{2{\Psi}}\right),
\eea
transforms as
\bea
\delta_\eps T = \eps \p T + 2 \p \eps T - \frac{1}{8\pi G} \p^3 \eps.
\eea
The charges associated with $\chi[\eps(\vec{\varphi})]$  then turn out to be
\bea
Q_\chi = \int_{\cal{H}} d\mathcal{H}\; T[\Psi]\, \eps,
\eea
where $d \mathcal{H}=\Gamma^{\frac{d-2}{2}}\sqrt{\det\gamma} d\theta d\vec{\varphi}$. {If} $Q_\chi$ for $\eps=e^{i\vec{m}\cdot\vec{\varphi}}$ is denoted by $L_{\vec{m}}$, the charge algebra $\{ Q_\chi,Q_{\chi^\prime} \}\equiv \delta_{\chi^\prime} Q_\chi $ exactly reproduces the NHEG algebra \eqref{NHEG-algebra}.

\vskip 2mm
\noindent\textbf{Discussion and outlook.}

\vskip 1mm

In this work we put forward a proposal for the semiclassical phase space of near-horizon extremal geometries which are solutions to vacuum Einstein gravity with $SL(2,\mathbb R)\times U(1)^{d-3}$ isometry. 
{We started with a solution of general relativity \eqref{NHEG-metric}, and showed that there is an infinite set of metrics \eqref{NHEG-phase-space} which, despite being diffeomorphic to each other, are physically distinct at the 
classical and semiclassical level, as they are labelled by the charges of the near-horizon generalized Virasoro symmetry algebra $\nhegalgebra$ \eqref{NHEG-algebra}, which  we derived. This algebra has the entropy as its central charge and carries most of the information about the background. In particular, $\vec{k}$ which measures the rate of change of the angular velocity at extremality with respect to the Hawking temperature, appears in its structure constants. Our analysis  may hence be viewed as  first steps toward a possible bottom-up construction of the extremal black hole microstates. 
}

Despite sharing the common goal of describing symmetries of extremal black holes using the covariant phase space formalism, our results have crucial conceptual and technical differences with the Kerr/CFT correspondence \cite{Kerr/CFT} and its variants and extensions \cite{Compere-review} in several respects: (i) Instead of specifying boundary conditions for metric perturbations, we specify the metric  perturbations everywhere in spacetime. Moreover, we are able to exponentiate these perturbations to build a smooth phase space;  (ii) Since our phase space admit a transitive action mapping any two points on $AdS_2$, surfaces charges are defined anywhere in the bulk, not only at infinity. The corresponding symmetries are therefore symplectic instead of asymptotic; (iii) Unlike the Kerr/CFT {proposal}, our symmetry algebra is not extension of  a $U(1)$ isometry of the background. Instead, it forms an additional direct product, \emph{cf.} \eqref{Agebra-Full}. All the points in the phase space are $SL(2,\mathbb R) \times U(1)^{d-3}$ invariant and the angular momenta are constant over the phase space; (iv) The choice of symmetry generator which preserves one null expansion-free congruence ($b=1$) allows us to build a smooth set of geometries, bypassing technical difficulties (conical defects, etc) of building a phase space for the choice of the Kerr/CFT generator ($b=0$) \cite{AMR}; (v) All the $U(1)$ directions appear democratically in our construction, both in the phase space and in the symmetry algebra. All expressions are  manifestly $SL(d-3,\mathbb{Z})$ covariant.

{The conserved charges labelling 
each geometry are built from an effective stress-tensor in terms of a field ${\Psi}$ on the torus $U(1)^{d-3}$ which we named the NHEG boson. This field provides a representation of the  $\nhegalgebra$ algebra \eqref{NHEG-algebra} which ressembles a $d-2$ dimensional version of the Liouville field theory. Such a theory is familiar for Einstein gravity in $AdS_3$ \cite{Coussaert:1995zp} but, to our knowledge, never appeared  in relationship with extremal black holes in four and higher dimensions. Interestingly, we note that the expression for the stress-tensor \eqref{T-tensor}  implies that the zero mode of the generalized Virasoro algebra $\nhegalgebra$, $L_{\vec{0}}$, is a positive definite operator over the semiclassical phase space and can hence be a good candidate for defining a Hamiltonian. We expect that in a fully quantized phase space, the algebra \eqref{NHEG-algebra} appears as the fundamental symmetry and the field theory based on ${\Psi}$ may appear as an effective description. It is of course very exciting to explore this direction which may be useful for a semiclassical microstate counting.
}

\begin{acknowledgments}

MMShJ would like to thank Hossein Yavartanoo for discussions at the early stages of this work. MMSHJ, AS and KH  would like to thank Allameh Tabatabaii Prize Grant of Boniad Melli Nokhbegan of Iran. MMSHJ, KH and AS would like to thank the ICTP network project NET-68. AS would like to thank the hospitality of ULB where this project was completed. G.C. is a Research Associate of the Fonds de la Recherche
Scientifique F.R.S.-FNRS (Belgium) and he acknowledges the current support of the ERC Starting Grant 335146 ``HoloBHC".

\end{acknowledgments}


\end{document}